\documentclass[Physsubmission, Phys]{SciPost}

\hypersetup{
    colorlinks,
    linkcolor={red!50!black},
    citecolor={blue!50!black},
    urlcolor={blue!80!black}
}

\usepackage[bitstream-charter]{mathdesign}
\DeclareSymbolFont{usualmathcal}{OMS}{cmsy}{m}{n}
\DeclareSymbolFontAlphabet{\mathcal}{usualmathcal}

\begin{document}

\def\bb    #1{\hbox{\boldmath${#1}$}}

\begin{center}{\Large \textbf{
Hadron Production in terms of Green's Functions in Non-Equilibrium Matter\\
}}\end{center}

\begin{center}
Andrew Koshelkin\textsuperscript{1},
\end{center}

\begin{center}
{\bf } National Research Nuclear University (MEPhI), Moscow, Russia
\\

*and.kosh59@gmail.com
\end{center}

\begin{center}
\today
\end{center}

\definecolor{palegray}{gray}{0.95}
\begin{center}
\colorbox{palegray}{
  \begin{tabular}{rr}
  \begin{minipage}{0.1\textwidth}
    \includegraphics[width=30mm]{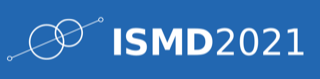}
  \end{minipage}
  &
  \begin{minipage}{0.75\textwidth}
    \begin{center}
    {\it 50th International Symposium on Multiparticle Dynamics}\\ {\it (ISMD2021)}\\
    {\it 12-16 July 2021} \\
    \doi{10.21468/SciPostPhysProc.?}\\
    \end{center}
  \end{minipage}
\end{tabular}
}
\end{center}

\section*{Abstract}
{\bf
Following  the quark-hadron duality concept,   we show that  the number of   hadrons generated in  the deconfinement  matter   is entirely  determined by the exact non-equilibrium Green's functions of partons in the medium  and the vertex function governing the probability of the confinement-deconfinement phase transition.   In such an approach,  compactifying  the standard (3+1) chromodynamics into  $ QCD_{xy} + QCD_{zt}$,  the rate of the hadrons produced in particle collisions   is derived in the explicit form provided that the hadronization  is the first order phase transition.   The pion production  is found to be in good agreement to the  experimental results on  the pion yield    in $pp$ collisions.}

\vspace{10pt}
\noindent\rule{\textwidth}{1pt}
\tableofcontents\thispagestyle{fancy}
\noindent\rule{\textwidth}{1pt}
\vspace{10pt}

\section{Introduction}
\label{sec:intro}
The hadronization of the deconfined matter arising in high-energy   particle collisions plays a key role in hadron production in $pp$, $pA$ and $AA$  collisions.  Such a problem is however  an extremely  complicate to be solved   in the unified  approach, starting from the fundamental theory of  QCD that leads to development of various models   describing the hadronization of the deconfined matter,  whose applicability depends  essentially on the energies of colliding particles. One of  them is the color flux tube approach based on the longitudinal dominance and transverse confinement.  The hadron production is found to be  the  leading process in the  hadron generation at the low and intermediate-$p_T$ region in high-energy $e^ + e^ -$ annihilations and $pp$ collisions [1-7], where the tube arises   between a quark and antiquark in  $e^+e^-$ annihilation reaction,  and between the valence quarks and antiquarks in  nucleon-nucleon collisions, respectively.

\section{Hadron production  in  quark-hadron duality concept}

The key assumption    we follow in the  derivation of hadron rate  is the concept of the quark-hadron duality\cite{Meln}. The central point of this concept is that  partons are the same particles in the confinement and deconfinement phases of the strong interaction matter. Then, the probability of the hadronization is proportional to the projection of the state vector of partons $| {q}_{deconf}>$ in the deconfinement matter on the such a vector determining the parton states $| {q}_{conf}>$ in the confinement medium

\begin{eqnarray}\label{eq81a}
&&{ \cal M} = < out | in>=<{\bar q}_{deconf}|{ q}_{conf}> ,
\end{eqnarray}
where $| {q}_{deconf}>$ and $| {q}_{conf}>$ mean the exact dressed quark states in the corresponding matters. 
When the transition of the quark states is governed by some unitarian operator $U$

\begin{eqnarray}\label{eq81aa}
&&|{ q}_{conf}> =U  | {q}_{deconf}> ,
\end{eqnarray}
then, the matrix element given by Eq.(\ref{eq81a}) is 

\begin{eqnarray}\label{eq81aaa}
&&{ \cal M} = <{\bar q}_{deconf}| U  | {q}_{deconf}> ,
\end{eqnarray}

When $| {q}_{deconf}>$ describes a single quark state the squared matrix element is expressed in terms of the single particle Green's function in a non-equilibrium medium $G^{-+} (x_1 , x_2)$\cite{LL-kin}

\begin{eqnarray}\label{eq81b}
&&<\vert { \cal M} \vert^2 >= - Tr \left\{  (U^\dag_2  \gamma^0 G^{-+}_{21}) (\gamma^0 U_1 G^{-+}_{1 2}) \right\} = Tr \left\{\vert  \gamma^0 U_1 G^{-+}_{1 2}) \vert^2 \right\},
\end{eqnarray}
where the trace symbol means summing with respect to all quantum number,  including integration over $x_1$ and $ x_2$, and averaging with respect to the deconfinement quark vacuum; the indexes denote  the coordinate $x$ and spin $\sigma$ variables, $1 = (\sigma , x )$. The subscribe at $U$ implies acting on the corresponding variable.

In the momentum representation we have the following 

\begin{eqnarray}\label{eq83bb}
&&\frac {d <\vert { \cal M} \vert^2 >}{dp }= - Tr \left\{ \int \frac{dq}{(2\pi)^8}
{\bar G}^{-+} (p +q/2)  ~ \varrho (p +q/2, p -q/2) ~ G^{-+} (p -q/2 ) \right\}.
\end{eqnarray}
where  $\varrho (p +q/2, p -q/2) ~ G^{-+} (p -q/2 )$  is the probability to  couple a quark-antiquark pair into a hadron whose  4-momentum is $p$.

Provided that the created hadron is on-shell , we obtain for the number of hadrons

\begin{eqnarray}\label{eq83bbbbb}
&&\frac{E({\bb p})dN_h} {{d^3 p} }  = \int d p^0 (E({\bb p}))\frac {d <\vert { \cal M} \vert^2 >}{dp },\end{eqnarray}
where $E({\bb p})=\sqrt{{\bb p}^2+m^2_h}$ and ${\bb p}$ are the hadron energy and momentum, respectively, $m_h$ is its mass.

\section{Hadron rate in longitudinal dominance approach}

The longitudinal  dominance and the transverse confinement in terms of the  QCD$_{xy}$+QCD$_{zt}$ dynamics\cite{Kos12} means the factorization  in Eq.(\ref{eq83bb}) with respect to the coordinate $x,y$ and $z,t$, so that the combined $p_T$ and rapidity distribution can be derived by overlapping  the Wigner functions corresponding to the parton motion in the transverse and longitudinal direction with the probability of hadron production. Then, going to  wave function in Eqs.(\ref{eq83bb}), (\ref{eq83bbbbb}), assuming the standard relation between them\cite{LL-kin} for both quarks and antiquarks,
we obtain

\begin{eqnarray}\label{5pp-rapt}
 \frac{dN_h} {{d^3 p} }=   \int \frac{d^3 \bb q}{(2\pi)^6}    \varrho ({\bb p} ; {\bb q})
\left | \Psi_{{\bar q}}  \left( {\bb p}_\bot + \frac{{\bb q}_\bot}{2} \right) \Psi_{{ q}}  \left( {\bb p}_\bot - \frac{{\bb q}_\bot}{2} \right) \Psi_{{\bar q}}  \left( { p}_z + \frac{{ q}_z}{2} \right) \Psi_{{ q}}  \left( { p}_z - \frac{{ q}_z }{2} \right)\right|^2,
\end{eqnarray}
where $\Psi_{q , {\bar q}} ({\bb p})$  is the wave function of a quark (antiquark) in the momentum representation.

The convolution with respect to ${\vec q}_\bot$ in Eq.(\ref{5pp-rapt}) is some function of ${\bb p}$ and $q_z$ which can be interpreted as the probability to create a hadron with the momentum ${\bb p}$, so that the difference in the momenta quark and antiquark, constituating  this hadron, is $q_z$ . Since  a boson is created due coupling  quarks with the same $q_z$, such a convolution is 

\begin{eqnarray}\label{convolution}
 {\rho} ({\bb p}, q_z) =    \int \frac{d^2 q_{\bot}}{(2\pi)^6}  E({\bb p}) \varrho({\bb p} ; {\bb q})
 | \Psi_{{\bar q}}  ( {\bb p}_\bot + {\bb q}_\bot /2 ) \Psi_{{ q}}  ( {\bb p}_\bot - {\bb q}_\bot /2 )|^2 \equiv  {\cal P} ({\bb p})\delta (q_z),
\end{eqnarray}
where $ {\cal P}_a ({\bb p})$ is a probability to create a boson with the momentum $\bb p$ which is assumed to be governed by the first order phase transition, so that

\begin{eqnarray}\label{prob-1st-k}
&& {\cal  P}({\bb p} , y)\propto \exp \left(-\frac {\Delta G^{} (E_t) - \Delta g^{} (T)}{T} \right), ~~~ \Delta g ^ {} (T)  = \alpha \ln \left( \frac{C^{}_{q_h} (T)}{C^{}_{q_q} (T)}\right)
\end{eqnarray}
where $\Delta G^{} $ is the change of the Gibbs thermodynamic potential due the  transition of particles from one phase to another, which is equal to the transverse  hadron energy $E_t$. The term $ \Delta g ^ {} (T) $, where $ C^{}_{q_h} (T) $ and $C^{}_{q_q}(T)$ are the quark concentrations in the deconfinement and confinement phases, guarantees correct behavior of the probability of the phase transition with respect to the transition temperature $T_c$. When temperature $T > T_c$ the concentration of quark    in the hadronic phase is equal to zero that leads to ${\cal P}_a  (\bb p_\bot , y) \equiv 0$ due to the logarithm. In the opposite case, when $T \leq T_c$ the number of quarks in the hadronic and deconfined phases are the same, so that the probability of the phase transition has the finite,  well-defined magnitude. 

As for the functions $ \Psi_{{ q, {\bar q}}}  \left( { p}_z \right)$ , they  are governed  by the $(1+1)$ Dirac equation\cite{Kos12} which fundamental solution, having been written in terms of the  kinematic rapidity $\eta$  and the proper time $\tau$\cite{Won94},   are in the massless case

\begin{eqnarray}\label{ld-8}
&&
\psi_\pm(\tau,\eta)= \frac{\exp \left( - \frac{ (\eta \mp\ln (\tau/\tau_0))^2}{2\sigma^2}\right)}{\sqrt{\sigma \pi^{1/2}}} , ~~~~\psi(\tau = \tau_0, \eta)=\frac{ \exp \left( - \frac{\eta^2}{2\sigma^2}\right)}{ \sqrt{\sigma\pi^{1/2}}},
\end{eqnarray}
where is assumed that at the initial time  $\tau=\tau_0+0$ the parton   has the rapidity localized near $\eta=0$, mainly inside the interval of the order of $\sigma$.
Substituting Eqs.(\ref{convolution})-(\ref{ld-8}) into Eq.(\ref{5pp-rapt}) we obtain  the hadron distribution  $d N_h/dy d^2 p_\bot$ with respect to the transverse momentum $p_\perp$ and rapidity $y$

\begin{eqnarray}\label{5pp-1-1}
 &&\frac{ d N_h} {{dy d^2 p_\bot}}= \frac{<N_{ch}(s)> \exp(m_h/T)   \theta (T_c -T)}{(2 \pi)^{3/2}    ( 1+( T/m_h )   ) \sum\limits_{a=1}^N\sigma_a^{-1} }
  \sum\limits_{a=1}^N \frac{\exp \left(-\frac { E_t }{T} \right)}{m_h T\sigma^2_a } \nonumber \\
  &&\Bigg\{\exp \left( - \frac{ 2(y -y_a)^2}{\sigma_a^2}\right)+ \exp \left( - \frac{2 (y +y_a)^2}{\sigma_a^2}\right)\Bigg\},
\end{eqnarray}
where $m_h$ is a hadron mass,   $T$  and $T_c$ are the matter  temperature and the phase transition temperature, respectively,  $\theta (z)$ is the unit step function,  $\sigma_a $ and $y_a$ are the parameter related to  the  initial beams of particles so that, $y_a = y_{beam} /2$ where $y_{beam}$ is the  rapidity of the initial beam which is  incident on a rest target. In this way $\sigma_ a =\sigma $ are fitted by a formula

\begin{eqnarray}\label{sigma-yb}
&&\frac{y_b}{2}= \left( \int \limits_{-\infty}^{+\infty} dy \vert \psi_{a\pm} (y)|^2\vert y^2 \right)^{1/2}= \frac{\sigma}{\sqrt 2},
\end{eqnarray}
whereas the obtained rate is assumed to be normalized by the  total hadron multiplicity  $ <N_{ch}(s)>$. 

The summation with respect to $a$ in the above formula  arises since in the LUND\cite{And83}, which we follow here, the  creation of hadron occurs due the fragmentation of the initial tube into secondary ones which are virtual hadron in the deconfinement matter, and hereby\cite{And83} $y=\eta$. The results of the comparison of the derived hadron distribution with the experimental results on pion production in $pp$ collisions are presented in Figs.1,2. 
\begin{figure}[!htb]
\begin{minipage}{0.38\textwidth}
     \centering
\includegraphics[width=5.5cm,angle=-90]{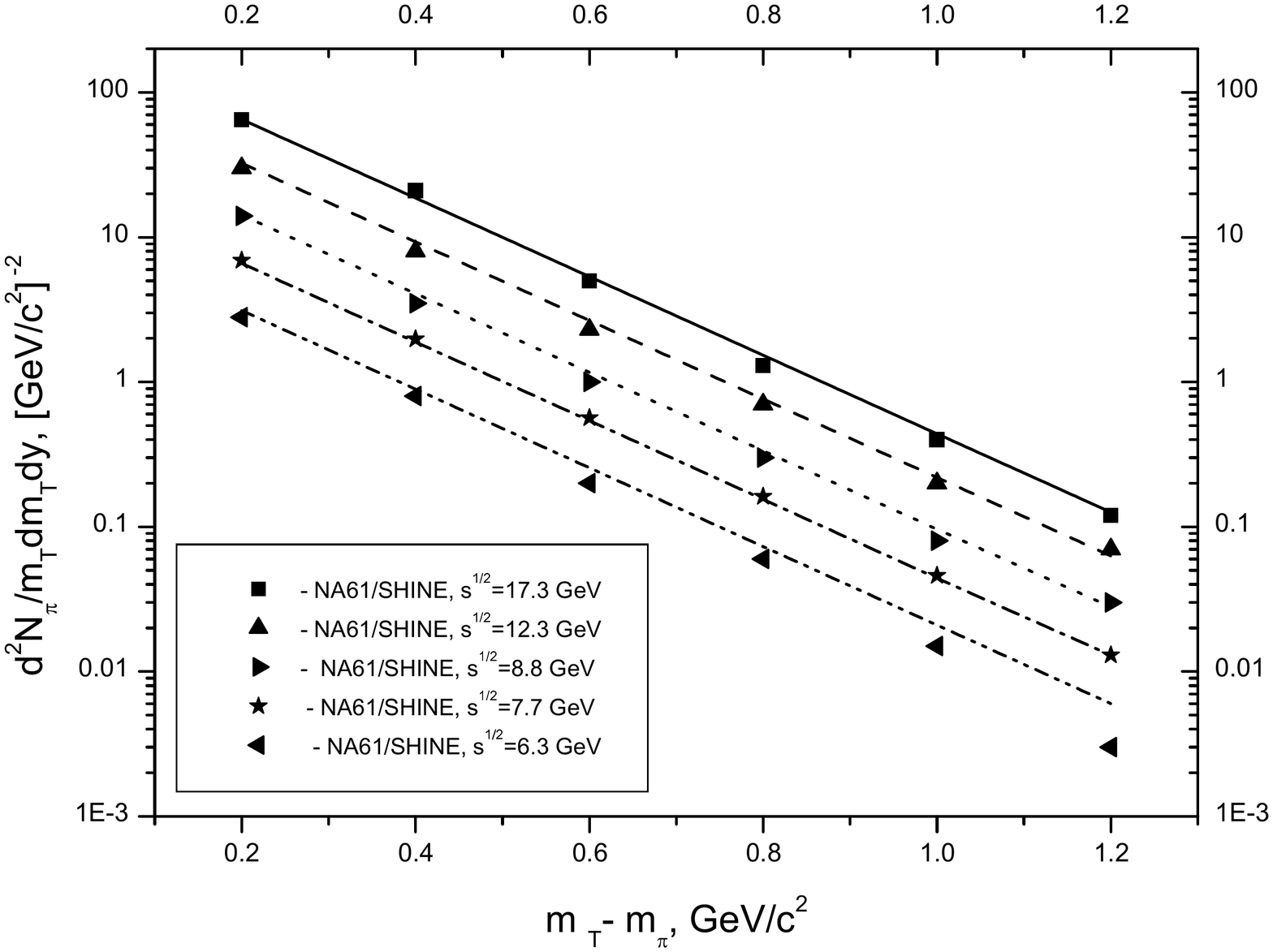}
\caption{\label{Fig.3} }
 \end{minipage}\hfill
 \begin{minipage}{0.48\textwidth}
     \centering
\includegraphics[width=5.5cm,angle=-90]{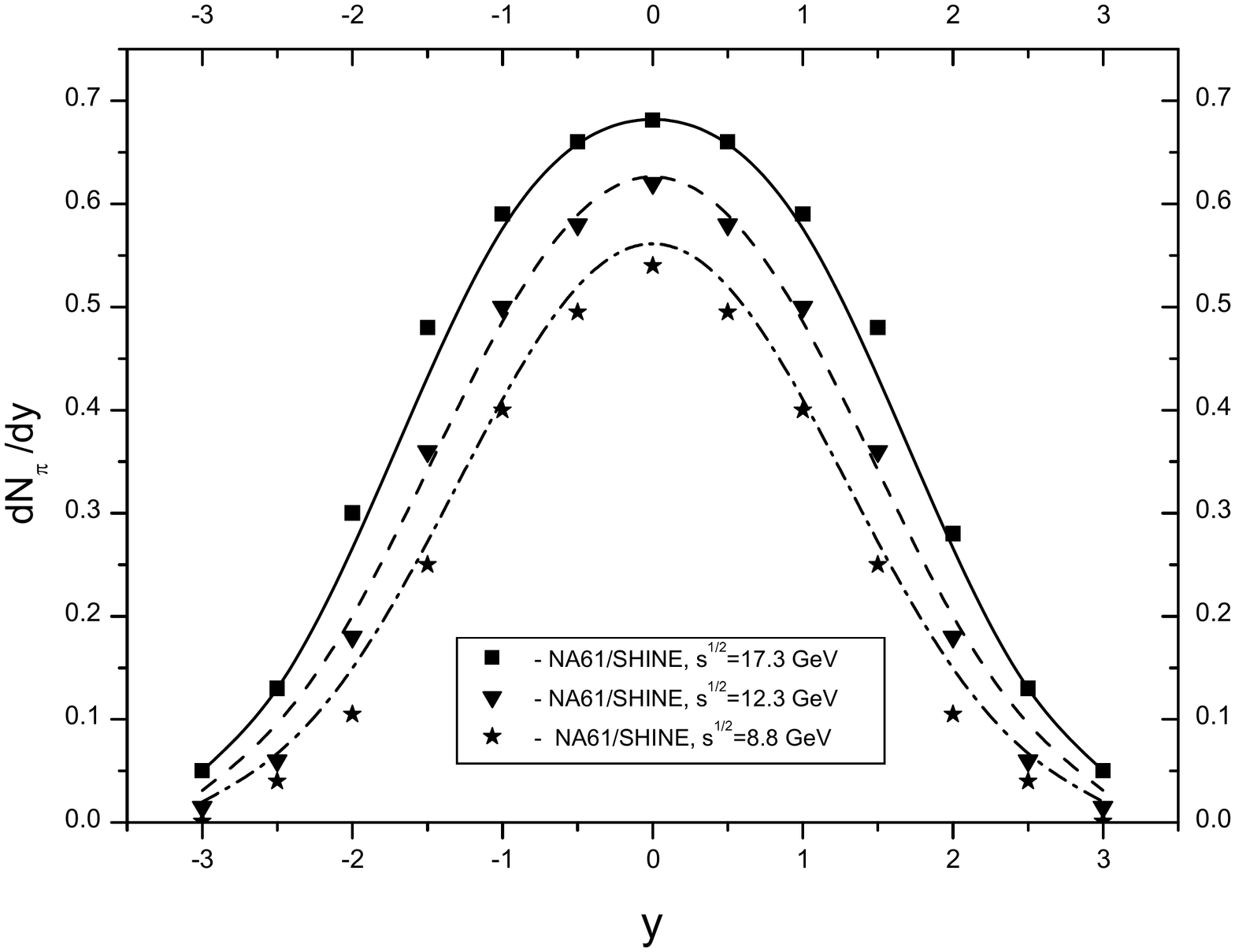}
\caption{\label{Fig.3} }
 \end{minipage}\hfill
\end{figure}
In  Fig.1 the lines of various types are the $p_{T}$ and rapidity distributions of pions  coming from  Eq.(\ref{5pp-1-1}) at $T_c =160 MeV$, and  are normalized by  the experimental value of the pion rate at $(m_{T} -m_\pi) = 0.2 GeV /c$, v.s. the pion rate  in $pp$ collisions\cite{NA61}(the scattered symbols) at the same projectile energies, where $m_{T} = \sqrt{p_T^2 + m^2_\pi}$,  whereas $m_\pi$ is the pion mass.
In Fig2. the rapidity distributions  coming from  Eq.(\ref{5pp-1-1}) at $\sqrt s = {17,3} GeV$ (solid lines), $\sqrt s = {12,3} GeV$ (dashed line), $\sqrt s = {8,8} GeV$ (dot-dashed line), and at $T_c= 160 MeV$ v.s. the rapidity distributions in $pp$ collisions\cite{NA61}.

\section{Conclusion}
Based on the quark-hadron duality concept\cite{Meln} we study  the hadronization of the deconfinement matter in the case of  a single quark-antiquark coupling. The hadron rate is found to be expressed in terms of the exact quark Green's functions in non-equilibrium matter and of the probability of the first-order equilibrium phase transition. Based on the   QCD$_{xy}$+QCD$_{zt}$ compactification \cite{Kos12} we derived both the  $p_T$ and rapidity distributions of  hadrons in the explicit form, provided that the hadronization is the equilibrium first-order phase transition.
When hadrons  are the pions generated  in the proton  collisions  of  intermediate energies, we have compared the hadron rate with the experimental results\cite{NA61}, and have gotten  to a good relation to  the experimental data for  all proton energies used in the experiment. We should note that the developed  approach should  be improved to get to a good relation to experimental data in the cases of p-A and A-A collisions that is in the  first turn connected with necessary to take into account of both multiple collisions of partons and  collective modes excited  in nuclei.

\nolinenumbers

\end{document}